\documentclass[twoside,twocolumn,english]{revtex4-1}
\usepackage[T1]{fontenc}
\usepackage[latin9]{inputenc}
\usepackage{amsmath}
\usepackage{amssymb}
\usepackage{graphicx}
\usepackage{esint}

\makeatletter

\usepackage{soul}
\usepackage{color}
\usepackage{babel}

\usepackage{babel}

\makeatother

\usepackage{babel}
\begin{document}

\title{Quantum Ultra-Walks: Walks on a Line with Hierarchical Spatial Heterogeneity}

\author{Stefan Boettcher}

\affiliation{Department of Physics, Emory University, Atlanta, GA 30322; USA }
\begin{abstract}
We discuss the model of a one-dimensional, discrete-time walk on a
line with spatial heterogeneity in the form of a variable set of ultrametric
barriers. Inspired by the homogeneous quantum walk on a line, we develop
a formalism by which the classical ultrametric random walk as well
as the quantum walk can be treated in parallel by using a ``coined''
walk with internal degrees of freedom. For the random walk, this amounts
to a $2^{{\rm nd}}$-order Markov process with a \emph{stochastic}
coin, better known as an (anti-)persistent walk. When this coin varies
spatially in the hierarchical manner of ``ultradiffusion,'' it reproduces
the well-known results of that model. The exact analysis employed
for obtaining the walk dimension $d_{w}$, based on the real-space
renormalization group (RG), proceeds virtually identical for the corresponding
quantum walk with a \emph{unitary} coin. However, while the classical
walk remains robustly diffusive ($d_{w}=\frac{1}{2}$) for a wide
range of barrier heights, unitarity provides for a quantum walk dimension
$d_{w}$ that varies continuously, for even the smallest amount of
heterogeneity, from ballistic spreading ($d_{w}=1$) in the homogeneous
limit to confinement ($d_{w}=\infty$) for diverging barriers. Yet,
for any $d_{w}<\infty$ the quantum ultra-walk never appears to localize. 
\end{abstract}
\maketitle

\section{Introduction\label{sec:Intro}}

Discrete-time quantum walks (QW)\ \cite{PortugalBook} have captured
the imagination of researchers in recent times. Their appreciation
started at least with the quadratic gain over any classical search
achieved by Grover's quantum search algorithm\ \cite{Gro97a}, with
QW as the main ``diffusing'' operation to spread information throughout
an idealized memory. This was followed by the realization that a \emph{coined}
QW\ \cite{SKW03,AKR05,Boettcher18a} in discrete time, distinctly
from continuous-time alternatives\ \cite{Childs04}, can bring such
a gain also to more realistic \emph{2d}-geometries (up to smaller,
logarithmic corrections to the leading $O(\sqrt{N})$ algorithmic
complexity in problem size $N$). Ever since, QW have become objects
of intense study\ \cite{Kempe03,VA12,PortugalBook}, beyond their
algorithmic interest, on various geometries, not least of all on the
one-dimensional line\ \cite{Ambainis01,konno_2003a,inui_2004a,bach_2004a,Inui05}.
In simple geometries like lattices\ \cite{MBSS02}, fractals\ \cite{Marquezino12,Patel12,Boettcher14b},
and hyper-cubes\ \cite{marquezino_2007a}, many of the features,
such as ballistic spreading\ \cite{Ambainis01,Inui05,Konno05} and
localization effects\ \cite{Inui05,schreiber_2011a,Falkner14a,Vakulchyk17,Mares19}
that distinguish QW from classical random walks (RW), are readily
analyzed mathematically. These studies have spawned experimentally
realizations to demonstrate transport and localization in QW\ \cite{Perets08,Peruzzo1500,Schreiber12,Sansoni12,Crespi13,Qiang2016,Ramasesh17}
and its search properties\ \cite{Figgatt2017,Foulger14,Tang18a,Tang18b}
that may become the foundation of future, controlled quantum computations.

The above-mentioned studies of solvable QW on the line are all based
on spatially homogeneous coins. Only few examples exist concerning
the dynamics of walks in heterogeneous but unitary environments, i.e.,
using coins that vary extensively with location quasi-periodically\ \cite{Shikano10}
or are drawn from some random ensemble\ \cite{Vakulchyk17}, each
yielding analytical insights only into localization properties in
some limits. Here, we discuss a \emph{1d}-walk in which coins possess
a strong spatial variation but with a hierarchical repetition of coins,
which we shall call the quantum ultra-walk. It is inspired by classical
models of diffusion over an ultrametric arrangement of barriers\ \cite{Ogielski85,Huberman85,Maritan86,Sibani86,Ceccatto87,Hoffmann88}
that was meant to describe ultra-slow relaxation. Among other things,
ordinary diffusion in such a hierarchy proved to be quite robust,
and the degree of heterogeneity had to reach a certain threshold before
a walk became sub-diffusive. 

Using a real-space renormalization group (RG) \cite{Pathria}, we
can analytically determine the walk dimension $d_{w}$\ \cite{Itzykson89},
characterizing the asymptotic scaling variable $x/t^{1/d_{w}}$ (or
pseudo-velocity\ \cite{Konno05}) for the walk, in closed form for
a parameter that determines the relative strength of barriers. For
example, $d_{w}$ describes the anomalous spread of the wave-function
with time in terms of the mean-square displacement, $\left\langle x^{2}\right\rangle \sim t^{2/d_{w}}$
. Classically, $d_{w}$ also determines how recurrent a walk is, i.e.,
if the spatial (or fractal) dimension $d_{f}$ is larger than $d_{w}$,
a walker might miss an arbitrarily close site forever, or might not
return to a previously visited site. This connection is known as Pólya's
recurrence theorem \cite{Polya1921}. So, the domain covered by such
a walk is quite porous while, in turn, that domain is almost certainly
compact for $d_{f}<d_{w}$. The RG has been previously employed to
obtain $d_{w}$ for QW with a homogeneous coin in various fractal
geometries\ \cite{Boettcher14b,Boettcher17a} and to elucidate the
complexity of Grover's search algorithm in terms of the \emph{spectral}
dimension $d_{s}$ of the search-space\ \cite{Boettcher18a}. After
a brief review of the RG for the homogeneous walk, we demonstrate
our procedure first by re-deriving the classical result in a novel
manner by using a $2^{{\rm nd}}$-order Markov process\ \cite{Renshaw94,Weiss94}.
It mimics the coined QW in all but the final step of the analysis
whilst using a stochastic instead of a unitary coin\ \cite{Boettcher13a}.
Despite of these parallels, quantum effects clearly assert themselves
in the final analysis and, thus, in the behavior obtained for $d_{w}$.

This paper is organized as follows: Sec.~\ref{sec:Backround} briefly
reviews the simple (homogeneous) walk on a line. In particular, we
introduce the dynamic equation that describes the evolution of the
discrete-time walk, classical or quantum, and its RG treatment. In
Sec.~\ref{sec:UltraWalkRG}, we develop the RG for the case of a
hierarchical dependence of such coins. In Sec.~\ref{sec:SolutionRW},
we choose a hierarchy of stochastic coins to derive the familiar classical
result, Eq.~(\ref{eq:dwPRW}). In Sec.~\ref{sec:SolutionQW}, we
then derive the solution for a corresponding hierarchy of unitary
coins. We conclude with a discussion of our results in Sec.~\ref{sec:Discussion}.

\section{Background on Walks\label{sec:Backround}}

\subsection{Analytic properties of the master equation\label{subsec:Analytic-Properties}}

The time evolution of walks are governed by the discrete-time master
equation\ \cite{Boettcher13a} 
\begin{equation}
\left|\Psi\left(t+1\right)\right\rangle ={\cal U}\left|\Psi\left(t\right)\right\rangle ,\label{eq:MasterEq}
\end{equation}
with propagator ${\cal U}$. This propagator is a stochastic operator
for a classical, dissipative RW. But in the quantum case it is unitary
and, thus, reversible. Then, in the discrete $N$-dimensional site-basis
$\left|x\right\rangle $ of some network, the probability density
function (PDF) is given by $\rho\left(x,t\right)=\psi_{x,t}=\left\langle x|\Psi\left(t\right)\right\rangle $
for RW, or by $\rho\left(x,t\right)=\left|\psi_{x,t}\right|^{2}$
for QW. 

Assuming that we possess the eigensolutions for the propagator, ${\cal U}\phi_{j}=u_{j}\phi_{j}$
with eigenvalues $u_{j}$ and an orthonormal set of eigenvectors $\phi_{j}\left(x\right)$,
then the formal solution of Eq.~(\ref{eq:MasterEq}) becomes $\psi_{x,t}=\sum_{j}a_{j}u_{j}^{t}\phi_{j}\left(x\right)$.
For a stochastic ${\cal U}$, aside from the unique ($+1$)-eigenvalue
of the stationary state, the remaining eigenvalues have $\left|u_{j}\right|<1$,
thus, according to Eq.~(\ref{eq:MasterEq}), the dynamics is uniquely
determined by $\rho(\vec{x},t)\sim e^{-t/\tau}$ for large times $t$
with $\tau=-1/\ln\max_{j}\left\{ \left|u_{j}\right|<1\right\} $.
In turn, for unitary ${\cal U}$, all eigenvalues are uni-modular,
$\left|u_{j}\right|=1$ for all $j$, such that $u_{j}=e^{i\theta_{j}}$
with real $\theta_{j}$. A discrete Laplace-transform (or ``generating
function'') \cite{Redner01} of the site amplitudes
\begin{equation}
\overline{\psi}_{x}\left(z\right)={\textstyle \sum_{t=0}^{\infty}}\psi_{x,t}z^{t}\label{eq:LaplaceT-1}
\end{equation}
has all its poles \textendash{} and hence those for $\overline{\rho}\left(x,z\right)$
\textendash{} located right on the unit-circle in the complex $z$-plane
\cite{Boettcher16},
\begin{equation}
\overline{\rho}\left(x,z\right)\propto\prod_{j,l}\left[1-z\,e^{i\left(\theta_{j}-\theta_{l}\right)}\right]^{-1}.\label{eq:rho_z}
\end{equation}
For the stochastic propagator, these poles are located at $z_{j}=1/u_{j}$,
accordingly, typically along the real-$z$ axis with $z_{j}>1.$These
facts will prove significant for the interpretation of the RG results
in Sec.~\ref{sec:SolutionQW}.

\subsection{Asymptotic scaling for walks\label{subsec:Asymptotic-Scaling-for}}

For RW, the probability density $\rho\left(\vec{x},t\right)$ to detect
a walk at time $t$ at site $\vec{x}$, a distance $x=\left|\vec{x}\right|$
from its origin, obeys the scaling collapse with the scaling variable
$x/t^{1/d_{w}}$, 
\begin{equation}
\rho\left(\vec{x},t\right)\sim t^{-\frac{d_{f}}{d_{w}}}f\left(x/t^{\frac{1}{d_{w}}}\right),\label{eq:collapse}
\end{equation}
where $d_{w}$ is the walk-dimension and $d_{f}$ is the fractal dimension
of the network \cite{Havlin87}. On a translationally invariant lattice
of any spatial dimension $d(=d_{f})$, it is easy to show that the
walk is always purely ``diffusive'', $d_{w}=2$, with a Gaussian
scaling function $f$, which is the content of many classic textbooks
on RW and diffusion \cite{Feller66I,Weiss94}. The scaling in Eq.~(\ref{eq:collapse})
still holds when translational invariance is broken or the network
is fractal (i.e., $d_{f}$ is non-integer). Such ``anomalous'' diffusion
with $d_{w}\not=2$ may arise in many transport processes \cite{Havlin87,Bouchaud90,Redner01}.
Thus, the determination of $d_{w}$ provides fundamental insights
into the physics of the spreading dynamics of a walk. 

For QW on ordinary lattices \cite{grimmett_2004a}, Eq.~(\ref{eq:collapse})
generically holds with $d_{w}=1$, indicating a ``ballistic'' spreading
of QW from its origin. This value has been obtained for various versions
of one- and higher-dimensional QW, for instance, with so-called weak-limit
theorems \cite{konno_2003a,grimmett_2004a,Segawa06,Konno08,VA12}.
The RG for discrete-time QW with a coin \cite{Boettcher13a,QWNComms13,Boettcher14b,Boettcher16}
was developed to expand the analytic tools to understand QW, say,
for networks that lack translational symmetries. This RG provides
\cite{Boettcher17a} principally similar results as in Eq.~(\ref{eq:collapse})
in terms of the asymptotic scaling variable $x/t^{1/d_{w}}$ (or pseudo-velocity
\cite{Konno05}), whose existence allows to collapse all data for
the probability density $\rho\left(\vec{x},t\right)$, aside from
oscillatory contributions (``weak limit''). 

\subsection{Coined walks on the line\label{sec:Master-Equations-for}}

For nearest-neighbor transitions on a line, the propagator referred
to in Eq.~(\ref{eq:MasterEq}) becomes 
\begin{equation}
{\cal U}={\textstyle \sum_{x}}A_{x}\left|x+1\right\rangle \!\!\left\langle x\right|+B_{x}\left|x-1\right\rangle \!\!\left\langle x\right|+M_{x}\left|x\right\rangle \!\!\left\langle x\right|,\label{eq:propagator}
\end{equation}
where $A_{x}$, $B_{x}$, and $M_{x}$ specify the (possibly position-dependent)
hopping operators for transitions to the left, right, or same site,
on leaving from a site $x$. RW merely requires local conservation
of probability, $A_{x}+B_{x}+M_{x}=1$, and could be satisfied with
a scalar Bernoulli coin $p$, such that $A=p$, $B=1-p$, and $M=0$,
for a homogeneous walk, for example. In contrast, conservation of
probability for $\rho\left(x,t\right)=\left|\psi_{x,t}\right|^{2}$
in QW demands unitary propagation, $\mathbb{I}={\cal U}^{\dagger}{\cal U}$,
which imposes on the ``coin-space'' the conditions $\mathbb{I}_{r}=A_{x}^{\dagger}A_{x}+B_{x}^{\dagger}B_{x}+M_{x}^{\dagger}M_{x}$,
$A_{x}^{\dagger}B_{x-2}=0$, and $0=A_{x}^{\dagger}M_{x+1}+M_{x}^{\dagger}B_{x+1}$.
For nontrivial choices satisfying $A_{x}^{\dagger}B_{x+2}=0$, this
algebra requires at least $r=2$-dimensional hopping matrices. It
is common\ \cite{PortugalBook,VA12} to construct the propagator
as a combination of a unitary $r\times r$ coin-matrix ${\cal C}_{x},$
acting on each site, and a shift operator ${\cal S}$. The coin mixes
the components of $\psi_{x,t}$ locally while the shift is represented
by a set of matrices $S^{(j)}$ connecting neighboring sites, providing
for a subsequent transfer in each direction. In this manner, coin
and spatial degrees of freedom become entangled.

For the propagator in Eq.~(\ref{eq:propagator}), this suggests the
choice of $A_{x}=S^{A}{\cal C}_{x}$, $B_{x}=S^{B}{\cal C}_{x}$,
and $M_{x}=S^{M}{\cal C}_{x}$ with $S^{A}+S^{B}+S^{M}=\mathbb{I}_{r}$.
The simplest form, $r=2$, the degree of each site on the line, does
not allow for self-loops ($S^{M}=0,M_{x}=0$) but shifts upper (lower)
components of each $\psi_{x,t}$ to the right (left) using the projectors
\begin{eqnarray}
S^{A}=\left[\begin{array}{cc}
1 & 0\\
0 & 0
\end{array}\right], & \qquad & S^{B}=\left[\begin{array}{cc}
0 & 0\\
0 & 1
\end{array}\right].\label{eq:Shift}
\end{eqnarray}
Then, it is easy to show that the unitarity conditions above are satisfied
for arbitrary unitary coins ${\cal C}_{x}$. 

\subsection{Renormalization of walks on a line\label{sub:Renormalization-of-walks:}}

The homogeneous RW or QW on a line, as defined above, is readily solved
to find $\rho(x,t)$ via a Fourier transform (see Ref.~\cite{PortugalBook},
for example). However, many geometries lack translational invariance,
in which case other methods need to be devised. One such method is
the real-space renormalization group (RG) \cite{Pathria} that is
particularly effective for hierarchical, recursively defined systems.
For the following, it is thus instructive to review the RG for a homogeneous
walk on the line, which is trivially recursive. 

To obtain the RG-recursions, the master equation (\ref{eq:MasterEq})
is projected into $\left|x\right\rangle $-space with ${\cal U}$
as given in Eq.~(\ref{eq:propagator}). Spatial homogeneity means
that the hopping operators are $x$-independent. Time is eliminated
by applying the generating function defined in Eq.~(\ref{eq:LaplaceT-1}),
such that the master equation becomes:
\begin{equation}
\overline{\psi}_{x}=zM\overline{\psi}_{x}+zA\overline{\psi}_{x-1}+zB\overline{\psi}_{x+1}.\label{eq:LMasternoX}
\end{equation}
For simplicity, we consider initial conditions (IC) localized at a
single site $x_{0}$, $\psi_{x,t=0}=\delta_{x,x_{0}}\psi_{IC}$. Eliminating
$\overline{\psi}_{x}$ for all sites for which $x$ is an odd number
and setting $x\to x/2$, the master equation reveals itself to be
\emph{self-similar} in form by appropriately redefining the renormalized
hopping operators $A$, $B$, $M$. To see this, we write for sites
adjacent to any even site $x$ \cite{Boettcher13a}: 
\begin{eqnarray}
\overline{\psi}_{x-1} & = & M\overline{\psi}_{x-1}+A\overline{\psi}_{x-2}+B\overline{\psi}_{x},\nonumber \\
\overline{\psi}_{x} & = & M\overline{\psi}_{x}+A\overline{\psi}_{x-1}+B\overline{\psi}_{x+1}+\delta_{x,x_{0}}\psi_{IC},\label{eq:1dPRWmass-master-1}\\
\overline{\psi}_{x+1} & = & M\overline{\psi}_{x+1}+A\overline{\psi}_{x}+B\overline{\psi}_{x+2}.\nonumber 
\end{eqnarray}
Solving this \emph{linear} system for the inner site $x$ yields $\overline{\psi}_{x}=M^{\prime}\overline{\psi}_{x}+A^{\prime}\overline{\psi}_{x-2}+B^{\prime}\overline{\psi}_{x+2}+\delta_{x,x_{0}}\psi_{IC}$,
leaving no effect on the IC, but requiring the (non-linear) RG recursions:
\begin{eqnarray}
A^{\prime} & = & A\left(\mathbb{I}-M\right)^{-1}A,\label{eq:recur1dPRWmass-1}\\
B^{\prime} & = & B\left(\mathbb{I}-M\right)^{-1}B,\\
M^{\prime} & = & M+A\left(\mathbb{I}-M\right)^{-1}B+B\left(\mathbb{I}-M\right)^{-1}A.\nonumber 
\end{eqnarray}
Physically, it expresses the effective behavior of a system in which
every other site had been coarse-grained out in terms of the renormalized
(primed) hopping operators, which now represent transitions over \emph{twice}
the distance of their (unprimed) priors. A recursive application of
Eq.~(\ref{eq:recur1dPRWmass-1}) reveals the asymptotic scaling of
the walk, as given by Eq.~(\ref{eq:collapse}), near the stationary
(``fixed'') points (FP) \cite{Redner01}. Linearizing the non-linear
system of RG recursions, such as Eq. (\ref{eq:recur1dPRWmass-1}),
around their FP provides a Jacobian matrix whose eigenvalues relate
the asymptotic behavior in space and time of the master equation in
Sec. \ref{subsec:Analytic-Properties}. While the eigenvalues of the
propagator determine the location of poles of $\overline{\rho}\left(x,z\right)$
in Eq. (\ref{eq:rho_z}), the Jacobian eigenvalues here determine
how these poles move in the complex $z$-plane under rescaling space.

In the classical analysis for RW with the scalar Bernoulli coin $p$
mentioned above, the recursions in Eq.~(\ref{eq:recur1dPRWmass-1})
yield for $z\to1$, i.e., $t\to\infty$ according to Eq.~(\ref{eq:LaplaceT-1}),
these three FP: $\left(A_{\infty},B_{\infty},M_{\infty}\right)=(0,0,M_{\infty})$,
$\left(1-M_{\infty},0,M_{\infty}\right)$, or $\left(0,1-M_{\infty},M_{\infty}\right)$,
where $M_{\infty}$ remains as an irrelevant constant that depends
on the details. Expanding Eq.~(\ref{eq:recur1dPRWmass-1}) to linear
order around each FP, the \emph{$2^{{\rm nd}}$} (\emph{$3^{{\rm rd}}$})
FP easily yields the ballistic solutions, corresponding to $d_{w}=1$
in Eq.~(\ref{eq:collapse}), that describes a drift to the left (right)
that is expected \emph{universally} for any $p<1/2$ ($p>1/2$). The\emph{
$1^{{\rm st}}$} FP is more delicate and can only be reached with
an unbiased coin, $p=1/2$, such that $A\equiv B\to0$ and $M\to M_{\infty}=1$
for $z\to1$. At this FP, a naive linearization fails. The self-term
$M$ dominates because, in unbiased diffusion, the ``range'' $L_{k}=2^{k}$
of a $k$-fold renormalized site outgrows the spread of RW at a time
$2^{k}$ such that almost all hops remain within that range, making
$M_{k}\sim1$. To subtract this trivial leading behavior, a correlated
solution has to be constructed with $A_{k}\equiv B_{k}\sim x_{k}\alpha^{k}$
and $M_{k}\sim1-y_{k}\alpha^{k}$ for large $k$ and $\left|\alpha\right|<1$.
Then, Eqs.~(\ref{eq:recur1dPRWmass-1}) yield $\alpha x_{k+1}=x_{k}^{2}/y_{k}$
and $\alpha y_{k+1}=y_{k}-2x_{k}^{2}/y_{k}$ with a single FP that
self-consistently determines $\frac{x_{\infty}}{y_{\infty}}=\alpha=\frac{1}{2}$.
The Jacobian $J_{k}=\frac{\partial\left(x_{k+1},y_{k+1}\right)}{\partial\left(x_{k},y_{k}\right)}$,
obtained from linearizing these recursions at its FP for $k\to\infty$,
gives $\lambda=4$ as the largest eigenvalue. Thus, for rescaling
size $L_{k+1}=2L_{k}$, time rescales as $t_{k+1}=\lambda t_{k}$,
as implied by the Tauberian theorems \cite{Weiss94,Hughes96,Redner01}.
Then, from $t_{k}\sim L_{k}^{d_{w}}$, we obtain $d_{w}=\log_{2}\lambda=2$
for the diffusive solution.

The corresponding asymptotic solution of QW with RG, in which case
Eq.~(\ref{eq:recur1dPRWmass-1}) is a set of \emph{matrix} recursions,
has been explored at length in Refs.~\cite{Boettcher13a,Boettcher17a}.
Since it will emerge as a special case of the discussion in Sec.~\ref{sec:SolutionQW},
we defer its consideration until then.

\section{Renormalization of Walks with an Ultrametric Set of Barriers\label{sec:UltraWalkRG}}

In the preceding section, we have reviewed how to use RG to solve
a homogeneous walk problem using a hierarchical approach. Having ignored
its translational invariance in our solution affords us the freedom
to explore more general, albeit strictly hierarchical problems on
the line. As the behavior of QW in heterogeneous environments is a
largely unexplored subject \cite{Shikano10}, investigating QW analytically
in a model with a spatially varying coin that respects the hierarchical
order seems to be a fruitful task. 

To that end, we consider position-dependent coins in such a way that
all sites of odd index $x$ share the same coin, and so do all sites
that are once-, twice-, trice-, $\ldots,i$-times divisible by 2.
We thus define the binary decomposition $x(i,j)=2^{i}(2j+1)$ with
a hierarchy-index, $i\geq0$, and running index, $-\infty<j<\infty$,
providing a unique, one-to-one relation between $x(\not=0)$ and the
pair $(i,j)$. Then, all sites $x\left(\not=0\right)$ that share
the same value of $i$ have an identical coin for all $j$, i.e.,
${\cal C}_{x(i,j)}={\cal C}_{i}$. 

In principle, there are many dynamically interesting sequences of
coins that could be defined on such a hierarchy. In the following,
we will focus on the case representing progressively more confining
barriers, which has been studied classically as a model for glassy
behavior. For instance, similar classical models have been proposed
for slow relaxation and aging\ \cite{Ogielski85,Huberman85,Maritan86,Sibani86,Ceccatto87,Hoffmann88}.
As even homogeneous QW have shown to exhibit peculiar localization
behavior \cite{Inui05,Falkner14a}, it is an interesting question
whether such barriers induce novel quantum effects that are absent
classically. 

\begin{figure}
\hfill{}\includegraphics[viewport=0bp 0bp 600bp 450bp,clip,width=1\columnwidth,viewport=0bp 0bp 630bp 430bp]{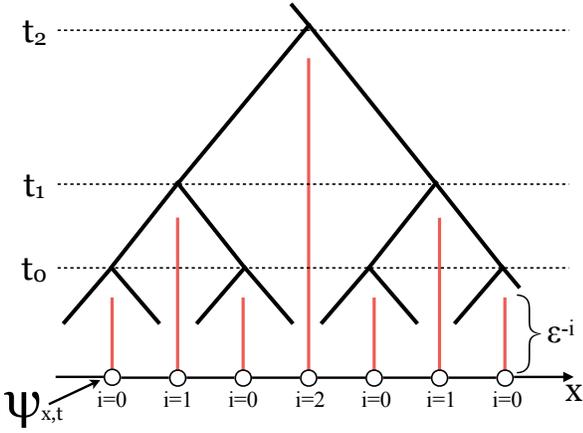}\hfill{}

\caption{\label{fig:UltraBarriers}Depiction of the hierarchical set of barriers
(red vertical lines) of relative reflectivity $\epsilon^{-i}$ for
$0<\epsilon<1$ and hierarchical index $i$ on a \emph{1d}-line, as
implemented in Eq.~(\ref{eq:eta_i}) for the classical walk and in
Eq.~(\ref{eq:eta_i}) for the quantum walk, generating an ultrametrically
arranged set of domains (illustrated by a tree) with a hierarchy of
characteristic timescales $t_{i}$ for escape.}
\end{figure}

A hierarchy of barriers arises when the sequence of such coins becomes
ever more \emph{reflective} for a walker trying to transition through
the respective site. Then, the walker gets confined in a tree-like
ultrametric set of domains with vastly varying timescales for exit.
Two neighboring domains at level $i$ form a larger domain at level
$i+1$, and so on, from which an ultrametric hierarchy emerges. Barriers
between such domains are depicted in Fig.~\ref{fig:UltraBarriers}.

\begin{figure}
\hfill{}\includegraphics[viewport=0bp 0bp 640bp 410bp,clip,width=0.85\columnwidth,viewport=10bp 20bp 630bp 400bp]{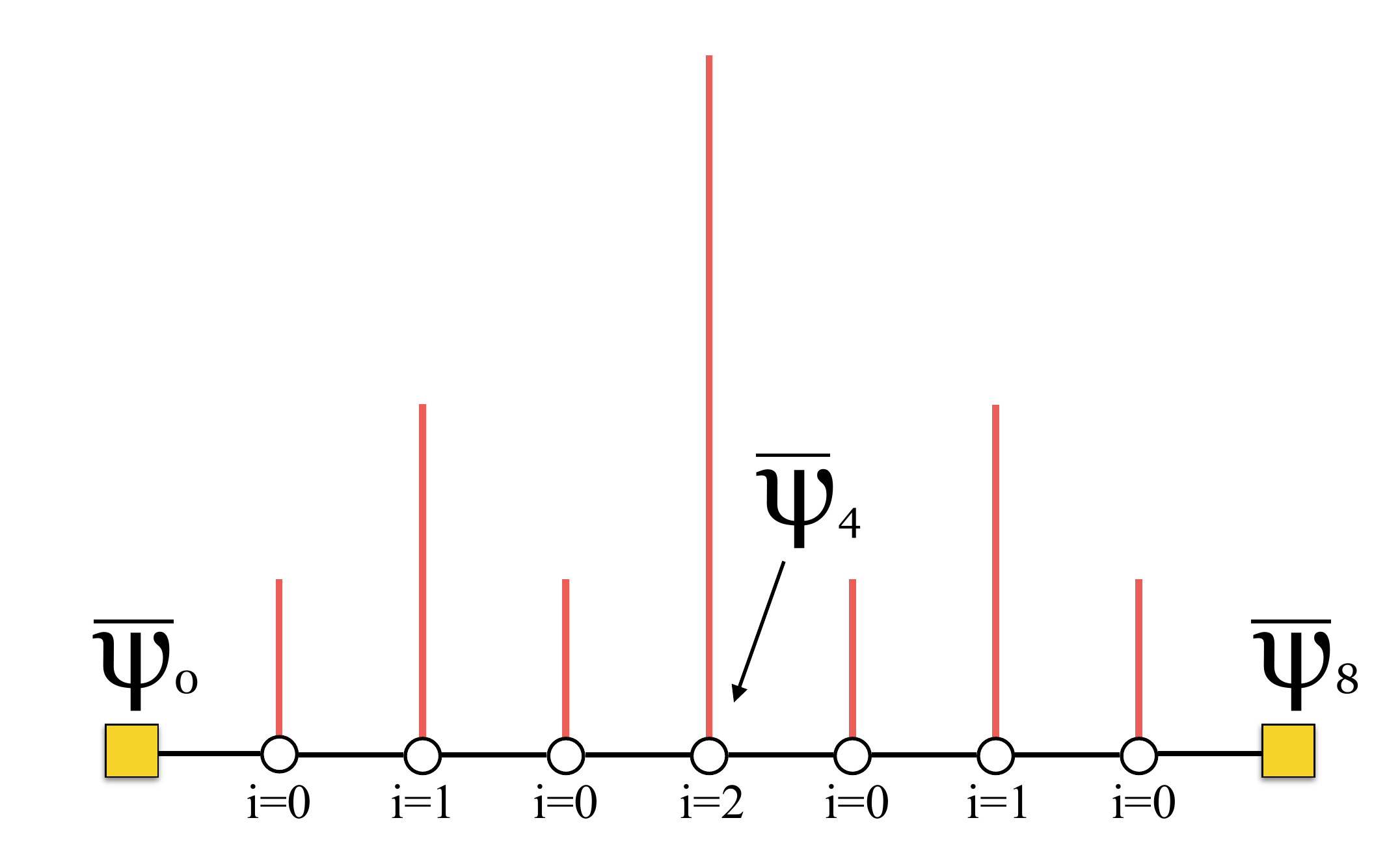}\hfill{}

\hfill{}\includegraphics[viewport=0bp 0bp 640bp 410bp,clip,width=0.85\columnwidth,viewport=10bp 20bp 630bp 400bp]{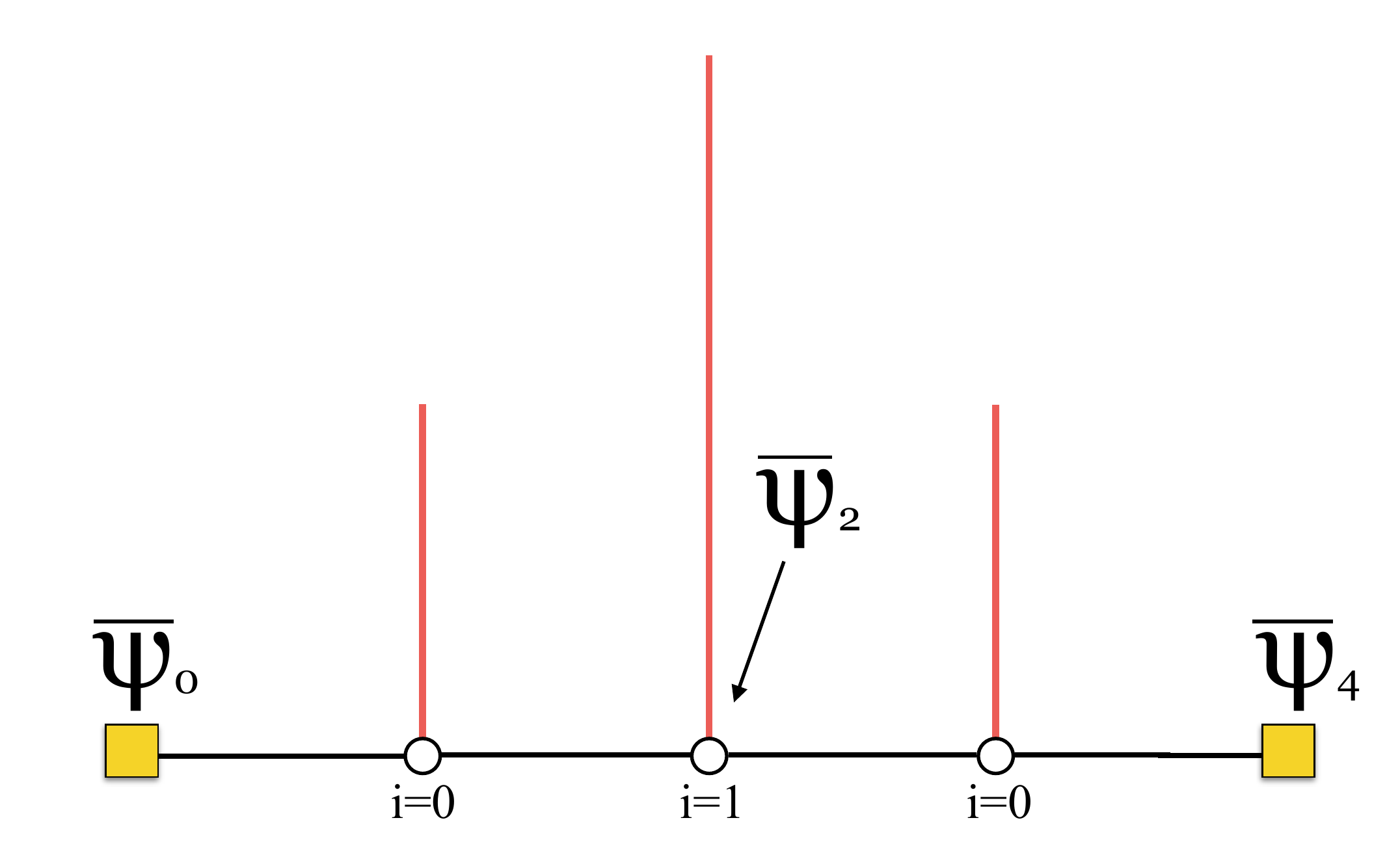}\hfill{}

\hfill{}\includegraphics[viewport=0bp 0bp 640bp 410bp,clip,width=0.85\columnwidth,viewport=10bp 20bp 630bp 400bp]{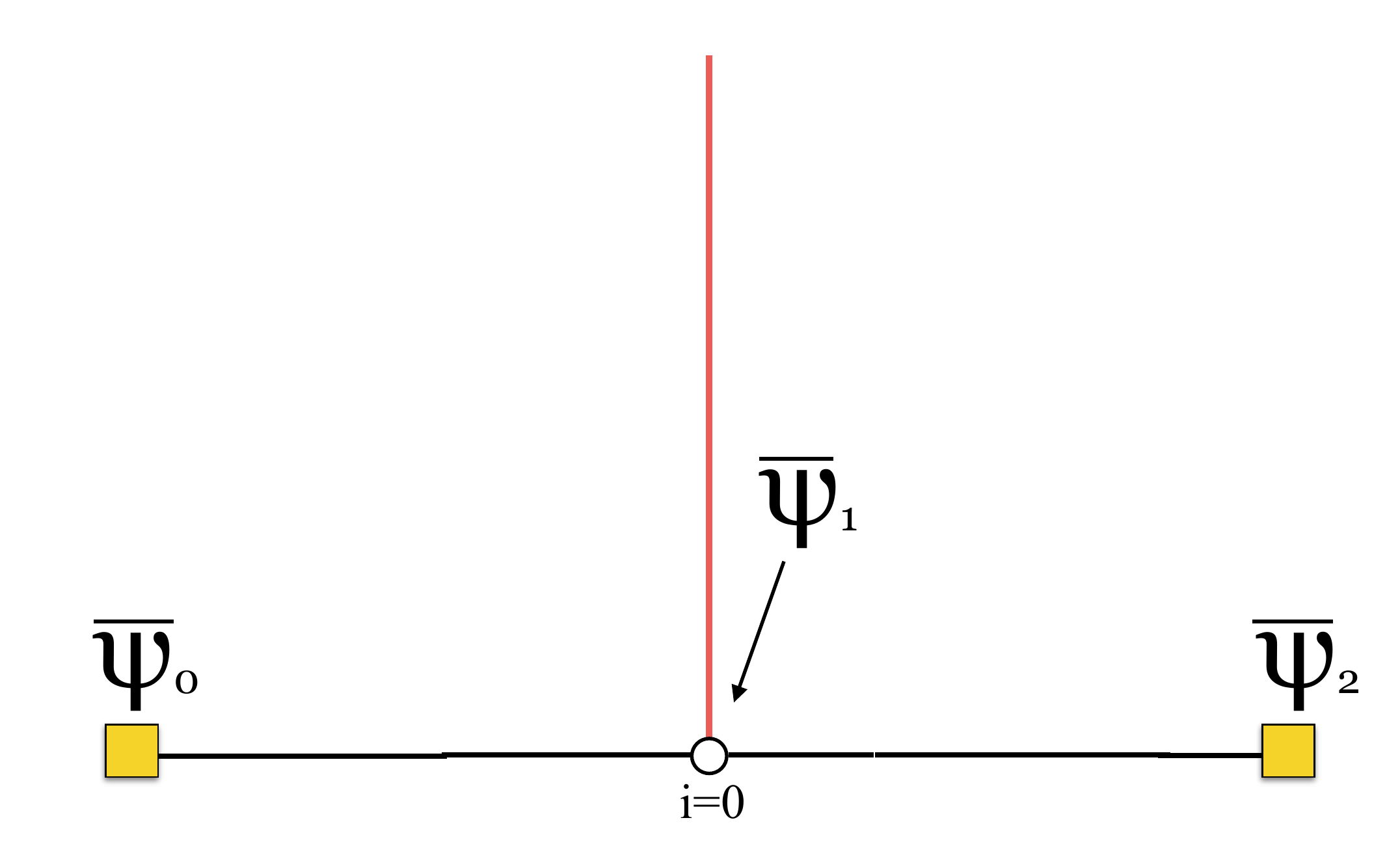}

\caption{\label{fig:UltraRG}Last three RG-steps $k=l-3$, $l-2$, and $l-1$,
in an ultra-walk with absorbing walls (yellow squares). Note that
for each RG-step $k\to k+1$, the remaining hierarchy indices $i$
for sites $x$ go from $i\to i-1$, as reflected in Eq.~(\ref{eq:recur1dPRWmass}).
Then, for all sites $x^{(k)}\to x^{(k+1)}=\frac{x^{(k)}}{2}$. The
final step $k=l-1\to l$ is described by Eqs.~(\ref{eq:AbsoWall}-\ref{eq:AbsoPsi}),
assuming that the walk started at the central site, $x=x^{(0)}=2^{l-1}=\frac{N-1}{2}$
in a finite system of size $N=2^{l}+1$, originally, i.e., at $x^{(l-1)}=1$.}
\end{figure}

It is straightforward to adapt the discussion of the RG for the homogeneous
case in Sec.~\ref{sub:Renormalization-of-walks:} to the propagator
in Eq.~(\ref{eq:propagator}) with position-dependent hopping operators.
This generalizes the master equation in Eq.~(\ref{eq:LMasternoX})
to 
\begin{equation}
\overline{\psi}_{x}=zM_{x}\overline{\psi}_{x}+zA_{x-1}\overline{\psi}_{x-1}+zB_{x+1}\overline{\psi}_{x+1}.\label{eq:LMasterX}
\end{equation}
 Again, we successively eliminate all sites for which $x$ is an odd
number ($i=0$) and set $x\to x/2$ ($i\to i-1$), thereby removing
an entire hierarchy with every iteration, each with an \emph{identical}
coin ${\cal C}_{i}$. Starting at $k=0$ with the ``raw'' hopping
operators $A_{i}^{(0)}=zA_{x(i,j)}$, $B_{i}^{(0)}=zB_{x(i,j)}$,
and $M_{i}^{(0)}=zM_{x(i,j)}\equiv0$, step-by-step for $k=0,1,2,\ldots$,
the master equation remains \emph{self-similar} in form by identifying
the renormalized hopping operators $A_{i}^{(k)}$, $B_{i}^{(k)}$,
$M_{i}^{(k)}$. In analogy to Eq.~(\ref{eq:1dPRWmass-master-1}),
we focus on a site $x=x(i_{+},j)$ with $i_{+}\geq2$, which pertains
to every \emph{fourth} site on the line. Note that all sites $\pm1$
or $\pm3$ hops removed from $x$ are of odd index ($i=0$), while
those $\pm2$ hops removed \emph{must} have $i=1$. At RG-step $k$
we have:\ 
\begin{eqnarray}
\overline{\psi}_{x-1} & = & M_{0}^{(k)}\overline{\psi}_{x-1}+A_{i_{+}}^{(k)}\overline{\psi}_{x}+B_{1}^{(k)}\overline{\psi}_{x-2},\nonumber \\
\overline{\psi}_{x} & = & M_{i_{+}}^{(k)}\overline{\psi}_{x}+A_{0}^{(k)}\overline{\psi}_{x+1}+B_{0}^{(k)}\overline{\psi}_{x-1},\nonumber \\
\overline{\psi}_{x+1} & = & M_{0}^{(k)}\overline{\psi}_{x+1}+A_{1}^{(k)}\overline{\psi}_{x+2}+B_{i_{+}}^{(k)}\overline{\psi}_{x},\label{eq:1dPRWmass-master}\\
\overline{\psi}_{x+2} & = & M_{1}^{(k)}\overline{\psi}_{x+2}+A_{0}^{(k)}\overline{\psi}_{x+3}+B_{0}^{(k)}\overline{\psi}_{x+1}.\nonumber 
\end{eqnarray}
Solving this \emph{linear} system for the even sites $x$, $x\pm2$,
$x\pm4,$ etc., yields 
\begin{eqnarray}
\overline{\psi}_{x} & = & M_{i_{+}-1}^{(k+1)}\overline{\psi}_{x}+A_{0}^{(k+1)}\overline{\psi}_{x+2}+B_{0}^{(k+1)}\overline{\psi}_{x-2},\nonumber \\
\overline{\psi}_{x+2} & = & M_{0}^{(k+1)}\overline{\psi}_{x+2}+A_{i_{+}-1}^{(k+1)}\overline{\psi}_{x+4}+B_{i_{+}-1}^{(k+1)}\overline{\psi}_{x},\label{eq:RGafter}
\end{eqnarray}
and so on. (We have ignored localized IC on some site, which have
no effect on the RG, similar to Sec.~\ref{sub:Renormalization-of-walks:}.)
Matching the solutions for those even sites to Eq.~(\ref{eq:RGafter}),
we can read off the RG-recursions for all $i>0$: 
\begin{eqnarray}
A_{i-1}^{(k+1)} & = & A_{0}^{(k)}\left[\mathbb{I}-M_{0}^{(k)}\right]^{-1}A_{i}^{(k)},\nonumber \\
B_{i-1}^{(k+1)} & = & B_{0}^{(k)}\left[\mathbb{I}-M_{0}^{(k)}\right]^{-1}B_{i}^{(k)},\label{eq:recur1dPRWmass}\\
M_{i-1}^{(k+1)} & = & M_{i}^{(k)}+A_{0}^{(k)}\left[\mathbb{I}-M_{0}^{(k)}\right]^{-1}B_{i}^{(k)}\nonumber \\
 &  & \quad+B_{0}^{(k)}\left[\mathbb{I}-M_{0}^{(k)}\right]^{-1}A_{i}^{(k)}.\nonumber 
\end{eqnarray}
Those RG-steps of decimation are illustrated in Fig.~\ref{fig:UltraRG}.
Note the close resemblance with Eq.~(\ref{eq:recur1dPRWmass-1})
above, to which Eq.~(\ref{eq:recur1dPRWmass}) reduces when the (lower)
hierarchy indices are removed and the (upper) indices $(k)$ and $(k+1)$
mark unprimed and primed operators, respectively.

Amazingly, we can simplify Eq.~(\ref{eq:recur1dPRWmass}) even further
and entirely eliminate the hierarchy-index $i$: If we define the
$k$-th renormalized shift matrices $S_{k}^{\{A,B,M\}}$ via 
\begin{eqnarray*}
\left\{ A,B,M\right\} _{i}^{(k)} & = & S_{k}^{\{A,B,M\}}{\cal C}_{i+k},
\end{eqnarray*}
which matches the definitions above for the un-renormalized systems
at $k=0$. In fact, at $k=0$, all shift matrices are given by Eq.~(\ref{eq:Shift}),
independent of position orhierarchy. Then, the $S_{k}^{\{A,B,M\}}$
remain $i$-independent for all $k\geq0$. When inserted into Eqs.~(\ref{eq:recur1dPRWmass}),
they satisfy the recursions, 
\begin{eqnarray}
S_{k+1}^{\{A,B\}} & = & S_{k}^{\{A,B\}}\left({\cal C}_{k}^{-1}-S_{k}^{M}\right)^{-1}S_{k}^{\{A,B\}},\nonumber \\
S_{k+1}^{M} & = & S_{k}^{M}+S_{k}^{A}\left({\cal C}_{k}^{-1}-S_{k}^{M}\right)^{-1}S_{k}^{B}\label{eq:Sflow}\\
 &  & \quad+S_{k}^{B}\left({\cal C}_{k}^{-1}-S_{k}^{M}\right)^{-1}S_{k}^{A},\nonumber 
\end{eqnarray}
which instead have an explicit $k$-dependence via the inverse coins
${\cal C}_{k}^{-1}$ of the $k$-th hierarchy \footnote{We assume that both, the unitary but also the stochastic coin, have
an inverse.}. 

Up to this point, there was no need to specify whether this ultra-walk
is classical or quantum. For RW, it would seem that we could simply
choose scalar Bernoulli coins $p_{x}$ again, such that $A_{x}=p_{x}$,
$B_{x}=1-p_{x}$, and $M_{x}=0$, say, to satisfy the local conservation
of probability, $A_{x}+B_{x}+M_{x}=1$. However, as the non-local
construction of Maritan and Stella\ \cite{Maritan86} illustrates,
it is not possible to obtain a walk model with ultrametric barriers
employing merely such plain scalar coins. Here, we present an alternative
version of that ultradiffusion model that is intuitive and has greater
conceptual simplicity at the expense of adding an internal (coin-)degree
of freedom. Such a construction\ \cite{Boettcher13a} is easily recognized
as a second-order Markov process\ \cite{Renshaw94} or a persistent
random walk (PRW)\ \cite{Weiss94}. It has the added benefit of being
\emph{completely analogous} to the above construction of QW: We merely
replace the unitary coin by a stochastic coin ${\cal C}_{x}$, then
the sum $A_{x}+B_{x}+M_{x}$ is also unitary or stochastic, respectively.

\subsection{Absorbing walls\label{subsec:Absorbing-Walls}}

While the spreading behavior in itself, characterized by the walk
dimension $d_{w}$, is the most fundamental property of a walk, other
physical properties may be of interest. Another interesting physical
quantity is the absorption of a walk at a wall, classically \cite{PhysRevE.51.3862,PhysRevLett.74.2410,PhysRevLett.75.3210}
or quantum \cite{Ambainis01}. In particular, since the spreading
behavior merely measures the dynamics of that part of the walk which
actually moves, determining the absorption of the walk at a confining
wall distant from the initial site provides information about how
much of the weight of the walk ever reaches the wall. In turn, if
that absorption does not become unity, some weight must have become
localized within a bounded domain within those walls. For classical
diffusion in a simply connected domain that would seem unphysical.
However, QW do exhibit such localization behavior, even in the absence
of disorder\ \cite{Inui05,schreiber_2011a,Falkner14a,Vakulchyk17,Mares19}.
Of course, absorption is merely an indirect measure of localization,
yet, sufficient to ascertain the lack of it.

As a specific situation for such a setting, it is most convenient
within the formalism we have developed to consider a walk between
two absorbing walls of separation $N=2^{l}+1$, equidistant from the
starting site $x_{0}=2^{l-1}$, as illustrated in Fig.~\ref{fig:UltraRG}.
Note that in this $1d$ geometry, these walls completely confine the
walk. As the wall-sites $x=0$ and $x=2^{l}$ are fully absorbing,
there is no flow out of those sites and at the end of $l-1$ RG-steps
the Eqs.~(\ref{eq:1dPRWmass-master}) reduce to 
\begin{eqnarray}
\overline{\psi}_{0} & = & A_{0}^{(l-1)}\overline{\psi}_{1},\nonumber \\
\overline{\psi}_{1} & = & M_{0}^{(l-1)}\overline{\psi}_{1}+\psi_{IC},\label{eq:AbsoWall}\\
\overline{\psi}_{2} & = & B_{0}^{(l-1)}\overline{\psi}_{1}.\nonumber 
\end{eqnarray}
Thus, for either wall it is 
\begin{equation}
\overline{\psi}_{\left\{ 0,2\right\} }=S_{l-1}^{\left\{ A,B\right\} }\left({\cal C}_{l-1}^{-1}-S_{l-1}^{M}\right)^{-1}\psi_{IC}.\label{eq:AbsoPsi}
\end{equation}
We will discuss below the RG-prediction for the absorption for both,
RW and QW.

\section{Solution of the Classical Ultra-Walk\label{sec:SolutionRW}}

Like for QW, we will find that the state variable describing PRW is
now a 2-component vector $\psi_{x,t}$, which here expresses a \emph{memory}
of the previous step. The physics of classical walks with such memory
(``persistence'') has been widely studied \cite{Weiss94,Renshaw94}.
Based on its prior behavior, the upper component $\psi_{x,t}^{+}$
refers to a walker with the preference to step to the right in the
next time-step, and the lower component $\psi_{x,t}^{-}$ indicates
a left-hop preference. The value of each component describes the probability
of finding a walker at that site $x$ and time $t$ in state ``$\pm$'',
and the total probability of finding a walker there, irrespective
of preference, is simply the sum of the two, $\rho_{x,t}=\psi_{x,t}^{+}+\psi_{x,t}^{-}$.
Ignoring a potential left-right bias here, we consider a walker coming
from the left (right) to have a probability $\eta_{x}$ to continue
to move right (left), and a probability $1-\eta_{x}$ to reverse direction
in the next step. For $\eta_{x}>\frac{1}{2}$ ($\eta_{x}<\frac{1}{2}$)
the walker exhibits (anti-)persistence, and for $\eta_{x}=\frac{1}{2}$
reduces again to an ordinary unbiased RW without memory. The master
equations then reads: 
\begin{eqnarray}
\overline{\psi}_{x}^{+} & = & z\eta_{x}\overline{\psi}_{x-1}^{+}+z\left(1-\eta_{x}\right)\overline{\psi}_{x-1}^{-}+\delta_{x,x_{0}}\overline{\psi}_{IC}^{+},\nonumber \\
\overline{\psi}_{x}^{-} & = & z\left(1-\eta_{x}\right)\overline{\psi}_{x+1}^{+}+z\eta_{x}\overline{\psi}_{x+1}^{-}+\delta_{x,x_{0}}\overline{\psi}_{IC}^{-},\label{eq:1dPRWplain}
\end{eqnarray}
where $\overline{\psi}_{IC}$ (with $\overline{\psi}_{IC}^{+}+\overline{\psi}_{IC}^{-}=1$)
represents the IC of PRW, which we place again at some site $x_{0}$.
Thus, $\overline{\psi}_{x}^{+}$ ($\overline{\psi}_{x}^{-}$) only
depends on hops from its left (right) neighbor; it is that inflow
which induces the ``$+$'' (``$-$'') state. We can then write
Eq.~(\ref{eq:1dPRWplain}) conveniently in matrix notation as a propagator
like Eq.~(\ref{eq:propagator}) with \cite{Boettcher13a}
\begin{equation}
A_{x}=\left(\begin{array}{cc}
\eta_{x} & 1-\eta_{x}\\
0 & 0
\end{array}\right),\qquad B_{x}=\left(\begin{array}{cc}
0 & 0\\
1-\eta_{x} & \eta_{x}
\end{array}\right),\label{eq:ABinit}
\end{equation}
and $M_{x}=0$. As for QW, we can decompose these matrices further
to write them as a combination of a shift and a coin matrix, $\{A,B,M\}_{x}=S^{\{A,B,M\}}{\cal C}_{x}$,
with the same shift matrices as in Eq.~(\ref{eq:Shift}). However,
here we introduce the \emph{stochastic} coin matrix 
\begin{equation}
{\cal C}_{x}=\left(\begin{array}{cc}
\eta_{x} & 1-\eta_{x}\\
1-\eta_{x} & \eta_{x}
\end{array}\right),\label{eq:PRW1dS}
\end{equation}
in which each row sums to unity.

\subsection{Ultradiffusion as hierarchically anti-persistent walk\label{subsec:Ultradiffusion-as-a}}

With the same choice of a hierarchically defined coin as in Sec.~\ref{sec:UltraWalkRG},
decomposing $x=x(i,j)$, PRW is renormalized exactly the same way
such as to obtain Eq.~(\ref{eq:Sflow}). {[}Note that ${\cal C}_{x}$
in Eq.~(\ref{eq:PRW1dS}) also has an inverse, except for $\eta_{x}=\frac{1}{2}$,
the degenerate case of an ordinary (non-persistent) walk, which we
can safely exclude in the following.{]} Then, it is easy to formulate
a simple PRW that is in the same universality class as the ultradiffusion
model solved in Ref.~\cite{Maritan86}, by choosing for $0\leq\epsilon\leq1$
and some $\eta_{0}(<\frac{1}{2}$, to ensure invertibility of all
coins): 
\begin{equation}
\eta_{x(i,j)}=\eta_{i}=\eta_{0}\epsilon^{i}.\label{eq:eta_i}
\end{equation}
For $\epsilon=1$, we expect to recover the ordinary PRW on a homogeneous
\emph{1d} lattice. Since $\eta_{i}<\frac{1}{2}$ and, indeed, rapidly
approaches zero, the walk is increasingly anti-persistent, i.e., ever-larger
domains form that are bordered by sites $x$ of high hierarchical
index $i$, frustrating the walk attempting to leave the domain with
an exponentially smaller probability (or higher barriers) with $i$.

Because the RG recursions in Eq.~(\ref{eq:Sflow}) is expressed in
terms of matrices, we need to find a parametrization of those recurring
shift-matrices in terms of scalar variables \cite{Boettcher13a}.
After several iterations of Eq.~(\ref{eq:Sflow}), starting from
the unrenormalized shift matrices in Eq.~(\ref{eq:Shift}), a pattern
soon appears that can be parametrized as\ 
\begin{equation}
S_{k}^{A}=\left(\begin{array}{cc}
a_{k} & 0\\
0 & 0
\end{array}\right),\,S_{k}^{B}=\left(\begin{array}{cc}
0 & 0\\
0 & a_{k}
\end{array}\right),\,S_{k}^{M}=\left(\begin{array}{cc}
0 & m_{k}\\
m_{k} & 0
\end{array}\right).\label{eq:1dPRWAnsatz}
\end{equation}
That pattern reproduces itself after a single iteration of Eq.~(\ref{eq:Sflow})
by replacing $k$ with $k+1$ and identifying
\begin{eqnarray}
a_{k+1} & = & \frac{\eta_{k}a_{k}^{2}}{\left(1-m_{k}\right)\left[1-(1-2\eta_{k})m_{k}\right]},\label{eq:1dPRWrecur1}\\
m_{k+1} & = & m_{k}+\frac{a_{k}^{2}\left[1-\eta_{k}-\left(1-2\eta_{k}\right)m_{k}\right]}{\left(1-m_{k}\right)\left[1-(1-2\eta_{k})m_{k}\right]},\nonumber 
\end{eqnarray}
as the RG flow, with initial conditions $a_{0}=z$ and $m_{0}=0$.
Note that this recursion is non-autonomous due to the explicit $k$-dependence
via $\eta_{k}$. We could supplement $\eta_{k}$ as a dynamical variable
via its recursion $\eta_{k+1}=\epsilon\eta_{k}$\ \cite{Maritan86}
and study fixed points of those three recursions. However, there is
a more elegant approach using the transformations 
\begin{eqnarray}
a_{k} & = & \frac{\eta_{k}}{1-2\eta_{k}}\alpha_{k},\label{eq:PRWtransform}\\
m_{k} & = & \frac{1-\eta_{k}}{1-2\eta_{k}}-\frac{\eta_{k}}{1-2\eta_{k}}\mu_{k},\nonumber 
\end{eqnarray}
which turn Eqs.~(\ref{eq:1dPRWrecur1}) into 
\begin{eqnarray}
\alpha_{k+1} & = & \frac{1}{\epsilon}\,\frac{\alpha_{k}^{2}}{\mu_{k}^{2}-1},\label{eq:PRWalphamu}\\
\mu_{k+1} & = & \left(1-\frac{1}{\epsilon}\right)+\frac{1}{\epsilon}\mu_{k}\left(1-\frac{\alpha_{k}^{2}}{\mu_{k}^{2}-1}\right),\nonumber 
\end{eqnarray}
where we have assumed $\frac{\eta_{k}\left(1-2\eta_{k+1}\right)}{\eta_{k+1}\left(1-2\eta_{k}\right)}\sim\frac{1}{\epsilon}$
and $\frac{1-\eta_{k+1}}{\eta_{k+1}}-\frac{\left(1-\eta_{k}\right)\left(1-2\eta_{k+1}\right)}{\eta_{k+1}\left(1-2\eta_{k}\right)}\sim1-\frac{1}{\epsilon}$,
which is justified for $\eta_{k}$ in Eq.~(\ref{eq:eta_i}) at large
$k$ for $\epsilon<1$, but trivially holds for the homogeneous walk
at $\epsilon=1$ also. Now, the RG-flow in Eq.~(\ref{eq:PRWalphamu})
is purely autonomous but depends non-trivially on the parameter $\epsilon$
that characterizes the strength of the ultrametric barriers. This
flow has two obvious fixed points, at $\alpha_{\infty}=\frac{1}{\epsilon}-2$
and $\mu_{\infty}=\frac{1}{\epsilon}-1$, and at $\alpha_{\infty}=\frac{1}{\epsilon}-\epsilon$
and $\mu_{\infty}=-\frac{1}{\epsilon}$. The second one can not be
reached by any physical initial condition of the flow, as $\mu_{\infty}<0$.
The first fixed point is physical for $0\leq\epsilon\leq\frac{1}{2}$,
where the largest eigenvalue of the Jacobian of the flow in Eq.~(\ref{eq:PRWalphamu})
is $\lambda=\frac{2}{\epsilon}$, which reproduces the anomalous walk
dimension, 
\begin{equation}
d_{w}=1-\log_{2}\epsilon,\label{eq:dwPRW}
\end{equation}
found in Ref.~\cite{Maritan86}. When $\epsilon\to\frac{1}{2}$,
$d_{w}\to2$ and the effect of the barriers becomes irrelevant such
that ordinary diffusion ensues for all $\frac{1}{2}\leq\epsilon\leq1$,
making diffusion rather robust against the introduction of such a
set of barriers. {[}In fact, for $\epsilon=1$ only, we can find a
closed form solution for all $k$ of the RG-flow in Eq.~(\ref{eq:PRWalphamu}),
as in Refs.~\cite{Boettcher17a,BoLi15}.{]} However, to find the
fixed point for the diffusive solutions for $\frac{1}{2}\leq\epsilon\leq1$
requires a scaling ansatz with $\alpha_{k}=\left(2\epsilon\right)^{-k}x_{k}$
and $\mu_{k}=1+\left(2\epsilon\right)^{-k}y_{k}$ such that 
\begin{eqnarray}
x_{k+1}\sim\frac{x_{k}^{2}}{y_{k}}, & \qquad & y_{k+1}\sim2y_{k}-\frac{x_{k}^{2}}{y_{k}},\label{eq:diffusionFP}
\end{eqnarray}
now independent of $\epsilon$, with fixed point $x_{\infty}=y_{\infty}$
and a Jacobian eigenvalue of $\lambda=4$, i.e., $d_{w}=2$. These
results are summarized in Fig.~\ref{fig:ultradw}.

Finally, we note that replacing the anti-persistent hierarchy of coins
with a persistent one, easily achieved by replacing $\eta_{x}\to1-\eta_{x}$
in Eq.~(\ref{eq:PRW1dS}), leads again to a purely diffusive walk
for all $\epsilon$. The asymptotic analysis for large $k$ with $\epsilon^{k}\to0$
now yields the $\epsilon\to1$ limit of Eq.~(\ref{eq:PRWalphamu})
even for $\epsilon<1$ that then results in Eq.~(\ref{eq:diffusionFP})
with $d_{w}=2$. It is easy to see that the physical situation of
a persistent hierarchy differs dramatically from the anti-persistent
one: For large $k$, all coins now become transmissive (i.e., the
identity matrix) instead of reflective, leaving mostly the non-diagonal
coins at all odd sites ($i=0$) to institute a simple persistent,
and effectively homogeneous, walk that remains in the same universality
class as ordinary diffusion\ \cite{Weiss94}.

\subsection{Classical walk with absorbing walls\label{subsec:RW-Abso}}

Evaluation of Eq.~(\ref{eq:AbsoPsi}) for the geometry of Fig.~\ref{fig:UltraRG}
using the coin in Eq.~(\ref{eq:PRW1dS}) and the RG parametrization
in Eq.~(\ref{eq:1dPRWAnsatz}) after $k=l$ RG-steps, then taking
$l\to\infty$, we readily obtain 
\begin{eqnarray}
\overline{\psi}_{\left\{ 0,2\right\} } & = & \frac{\alpha_{\infty}}{\mu_{\infty}^{2}-1}S^{\left\{ A,B\right\} }\left(\begin{array}{cc}
1 & \mu_{\infty}\\
\mu_{\infty} & 1
\end{array}\right)\psi_{IC},\nonumber \\
 & = & S^{\left\{ A,B\right\} }\left(\begin{array}{cc}
\epsilon & 1-\epsilon\\
1-\epsilon & \epsilon
\end{array}\right)\psi_{IC},\label{eq:classAbso02}
\end{eqnarray}
with $S^{\left\{ A,B\right\} }$ given in Eq.~(\ref{eq:Shift}).
In PRW, the norm of a state-vector is simply the sum of its (certainly
non-negative) components, $\left\Vert \psi_{x,t}\right\Vert \stackrel{def}{=}\left(\begin{array}{c}
1\\
1
\end{array}\right)\circ\psi_{x,t}$. Then, for an absorbing site $x$, the absorption there is $F_{x}=\sum_{t=0}^{\infty}\left|\psi_{x,t}\right|=\left\Vert \overline{\psi}_{x}\left(z=1\right)\right\Vert $.
With $S^{A}+S^{B}=\mathbb{I}$, we finally get for the total (combined)
absorption:

\begin{equation}
F_{0+2}=\left\Vert \overline{\psi}_{0}+\overline{\psi}_{2}\right\Vert =\left\Vert \psi_{IC}\right\Vert =1,
\end{equation}
since $\psi_{IC}$ is normed to unity, of course. Thus, for any size
barrier $\epsilon$ and any system size, any walk started in the middle
will eventually get absorbed with certainty! For a classical walk
on the 1d-line, we would have expected this, due to Polya's theorem
\cite{Polya1921}, even for a heterogeneous environment.

\section{Solution of the Quantum Ultra-Walk\label{sec:SolutionQW}}

To design a quantum analogue to the classical PRW on an ultrametric
set of barriers, specified by Eq.~(\ref{eq:PRW1dS}), we consider
the real, unitary quantum coins 
\begin{equation}
{\cal C}_{i}=\left(\begin{array}{cc}
\sin\eta_{i} & \cos\eta_{i}\\
\cos\eta_{i} & -\sin\eta_{i}
\end{array}\right),\qquad\eta_{i}=\eta_{0}\epsilon^{i}\quad(0<\epsilon\leq1),\label{eq:etaCoin}
\end{equation}
although many other interesting choices may exist\ \cite{Vakulchyk17}.
Note that for $\epsilon=1$, this reproduces a homogeneous \emph{1d}
QW\ \cite{Boettcher13a}. However, for $\epsilon<1$, the coins become
increasingly resistant to transit, with $\eta_{i}\to0$ for $i\to\infty$,
blocking the transition through sites $x$ of higher index $i$, no
matter from which direction those sites are approached.

As for the classical case, evolving the recursions in Eq.~(\ref{eq:Sflow})
with this coin for a few iterations from the unrenormalized values,
already after one iteration a recurring pattern emerges that suggest
the Ansatz 
\begin{equation}
S_{k}^{A}=\left(\begin{array}{cc}
a_{k} & 0\\
0 & 0
\end{array}\right),\,S_{k}^{B}=\left(\begin{array}{cc}
0 & 0\\
0 & -a_{k}
\end{array}\right),\,S_{k}^{M}=\left(\begin{array}{cc}
0 & m_{k}\\
m_{k} & 0
\end{array}\right),\label{eq:QWparam}
\end{equation}
amazingly similar to the classical case in Eq.~(\ref{eq:1dPRWAnsatz}).
When iterated, the RG-recursions in Eq.~(\ref{eq:Sflow}) for the
scalar parametrization with $a_{k}$ and $m_{k}$ closes after each
iteration for 
\begin{eqnarray}
a_{k+1} & = & \frac{a_{k}^{2}\sin\eta_{k}}{1-2m_{k}\cos\eta_{k}+m_{k}^{2}},\label{eq:1dQWrecursions}\\
m_{k+1} & = & m_{k}+\frac{\left(m_{k}-\cos\eta_{k}\right)a_{k}^{2}}{1-2m_{k}\cos\eta_{k}+m_{k}^{2}},\nonumber 
\end{eqnarray}
with $a_{k=1}=z^{2}\sin\eta_{0}$, $m_{k=1}=z^{2}\cos\eta_{0}$, and
$\eta_{1}=\eta_{0}\epsilon$ as initial conditions. {[}Only the first
step, from $k=0$ to $k=1$, does not fit this pattern.{]} Note the
striking similarity of these recursions to those for PRW in PRW in
Eq.~(\ref{eq:1dPRWrecur1}). Like those, Eq.~(\ref{eq:1dQWrecursions})
is non-autonomous. Analogous to Eq.~(\ref{eq:PRWtransform}), we
can find an elegant transformation, 
\begin{eqnarray}
a_{k}=\alpha_{k}\sin\eta_{k}, & \qquad & m_{k}=\cos\eta_{k}-\mu_{k}\sin\eta_{k},\label{eq:UQWtransform}
\end{eqnarray}
that turns Eq.~(\ref{eq:1dQWrecursions}) into 
\begin{eqnarray}
\alpha_{k+1} & \sim & \frac{1}{\epsilon}\,\frac{\alpha_{k}^{2}}{\mu_{k}^{2}+1},\label{eq:1dQWrecIntermediate}\\
\mu_{k+1} & \sim & \frac{1}{\epsilon}\,\mu_{k}\,\left(1+\frac{\alpha_{k}^{2}}{\mu_{k}^{2}+1}\right)+\frac{1}{2}\eta_{k}\left(\epsilon-\frac{1}{\epsilon}\right),\nonumber 
\end{eqnarray}
where we have approximated $\frac{\sin\eta_{k+1}}{\sin\eta_{k}}\sim\epsilon$
and $\left(\cos\eta_{k}-\cos\eta_{k+1}\right)/\sin\eta_{k+1}\sim\frac{1}{2}\eta_{k}\left(\epsilon-1/\epsilon\right)$,
to within exponentially small corrections in $k$ for $\epsilon<1$,
and trivially correct for $\epsilon=1$.

\begin{figure}
\hfill{}\includegraphics[clip,width=1\columnwidth,viewport=10bp 30bp 750bp 540bp]{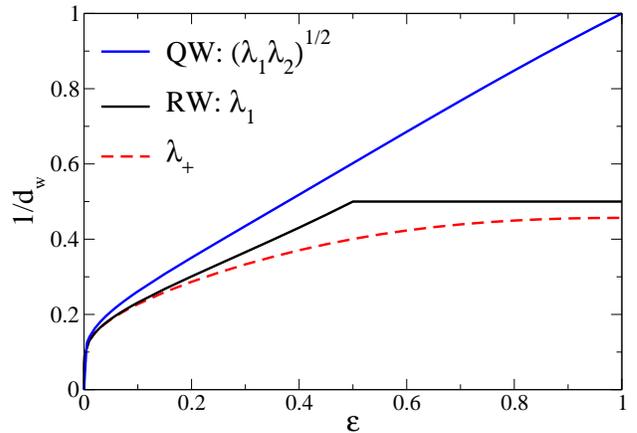}\hfill{}\caption{\label{fig:ultradw}Plot of the inverse walk dimension $1/d_{w}$
for the classical (RW) and the quantum walk (QW) with ultrametric
barriers as a function of coin-parameter $\epsilon$, where smaller
$\epsilon$ represents higher barriers between an ultrametrically
arrange hierarchy of domains that confine the walk. The black and
the blue lines are the respective RG-predictions in Eq.~(\ref{eq:dwPRW})
and in Eq.~(\ref{eq:dwQ}). Also, the unphysical prediction from
the eigenvalue $\lambda_{+}$ in Eq.~(\ref{eq:lambdaplus}) is shown
as red-dashed line. }
\end{figure}

If we were to consider neglecting the last term in Eq.~(\ref{eq:1dQWrecIntermediate}),
that is exponentially small for $\epsilon<1$, we find a fixed point
at $\alpha_{\infty}=1-\frac{1}{\epsilon}$ and $\mu_{\infty}=\pm i\sqrt{1-\frac{1}{\epsilon}+\frac{1}{\epsilon^{2}}}$.
Having $\alpha_{\infty}<0$ imposes no restriction, since the definition
of $a_{k}$ in Eq.~(\ref{eq:QWparam}) is invariant under $a_{k}\to-a_{k}$.
Having imaginary $\mu_{\infty}$ could be expected as a small correction
off the real axis to $m_{\infty}=1$ in a quantum problem. The associated
eigenvalues are very interesting, with 
\begin{eqnarray}
\lambda_{\pm} & = & \left(\frac{1}{\epsilon}+\frac{1}{2}+\epsilon\right)\pm\sqrt{\left(\frac{1}{\epsilon}+\frac{1}{2}+\epsilon\right)^{2}-2},\label{eq:lambdaplus}
\end{eqnarray}
of which only $\lambda_{+}>1$ for $0<\epsilon<1$. In fact, it reproduces
the eigenvalues found for a corresponding tight-binding spectrum considered
in Ref.~\cite{Ceccatto87}. As shown in Fig.~\ref{fig:ultradw},
it meets up with the classical result for $d_{w}$ really well for
$\epsilon\to0$. However, it generally predicts a slower spread than
even the classical walk throughout. Closer inspection of the boundary
layer\ \cite{BO} at $\epsilon\to1$ shows that it is actually a
very subtle extension of the (subdominant) diffusive solution found
for the ordinary \emph{1d} QW discussed in Ref.~\cite{Boettcher13a}.
While $\lambda_{+}\to\left(5+\sqrt{17}\right)/2\approx4.56...$ for
$\epsilon\to1$, this fixed point is actually not valid for $\epsilon=1$,
since $\mu_{\infty}\to\pm i$ and $\alpha_{\infty}\to0$, for which
the recursions in Eq.~(\ref{eq:1dQWrecIntermediate}) are singular.
Resolving that singularity with a scaling ansatz reproduces the diffusive
solution with $\lambda=4$ to which $\lambda_{+}$ discontinuously
connects.

To reveal the physically relevant scaling of QW, we argue as follows:
Without the last term in Eq.~(\ref{eq:1dQWrecIntermediate}), there
is also a fixed point values of $\alpha_{\infty}=\epsilon$ and $\mu_{\infty}=0$,
which in turn is inconsistent with dropping even an exponentially
small term, however. Rather, we retain the term and apply the Ansatz
$\mu_{k}=\eta_{k}\nu_{k}\ll1$ with $\nu_{k}$ finite to turn all
of Eq.~(\ref{eq:1dQWrecIntermediate}) autonomous:

\begin{eqnarray}
\alpha_{k+1} & \sim & \frac{\alpha_{k}^{2}}{\epsilon},\label{eq:UQWflow}\\
\nu_{k+1} & \sim & \frac{\nu_{k}}{\epsilon^{2}}\,\left(1+\alpha_{k}^{2}\right)-\frac{1-\epsilon^{2}}{2\epsilon^{2}},\nonumber 
\end{eqnarray}
The flow in Eq.~(\ref{eq:UQWflow}) has a fixed point at $\alpha_{\infty}=\epsilon$
and $\nu_{\infty}=\frac{1}{2}\left(1-\epsilon^{2}\right)$. Its Jacobian
eigenvalues are $\lambda_{1}=1+\epsilon^{-2}$ and $\lambda_{2}=2$,
i.e., $\lambda_{1}\geq\lambda_{2}>1$ for $0\leq\epsilon\leq1$. For
a classical walk with a stochastic master equation, the leading eigenvalue
$\lambda_{1}$ suffices to describe the walk dimension $d_{w}=\log_{2}\lambda_{1}$\ \cite{Redner01}.
However, it has been shown\ \cite{Boettcher16,Boettcher17a} that
the unitarity constraint imposed on the master equation for QW also
requires physical observables to be unitary. . Yet, the renormalized
parameters $a_{k}$ and $m_{k}$ are not, and their poles move with
$k$ both, tangentially and radially, to the unit circle in the complex
$z$-plane, while the poles of actual observables \emph{only} move
tangentially on the circle, see Eq.~(\ref{eq:rho_z}). Detailed analysis
shows\ \cite{Boettcher16} that the largest,dominant eigenvalue,
$\lambda_{1},$ only describes the tangential movement of poles and,
thus, must be ignored \footnote{In the appendix of Ref.~\ \cite{Boettcher16} an example was provided
that shows by explicit construction how the poles of the non-unitary
hopping parameters cancel as soon as these parameters are combined
to calculate an observable.}. Instead, the tangential movement of poles is found to be described
by the geometric mean of first and second eigenvalue, $\sqrt{\lambda_{1}\lambda_{2}}$.
Accordingly, we conclude that 
\begin{equation}
d_{w}^{Q}=\log_{2}\sqrt{\lambda_{1}\lambda_{2}}=\frac{1}{2}+\frac{1}{2}\log_{2}\left(1+\epsilon^{-2}\right).\label{eq:dwQ}
\end{equation}
This walk dimension, also shown in Fig.~\ref{fig:ultradw}, has all
the characteristics of being physical, as it is ballistic ($d_{w}=1$)
for the homogeneous \emph{1d} QW at $\epsilon\to1$, and it diverges
for diverging barrier heights, $\epsilon\to0$, where to leading order
matches up with the classical result, $d_{w}\sim-\log_{2}\epsilon$.
It also predicts the fastest spread for QW compared to RW or the fixed
point leading to Eq.~(\ref{eq:lambdaplus}). Note also that even
a minute disturbance of homogeneity, i.e., barriers for arbitrarily
small $1-\epsilon$, QW ceases to be ballistic.

\begin{figure}
\hfill{}\includegraphics[viewport=0bp 0bp 700bp 600bp,clip,width=1.2\columnwidth]{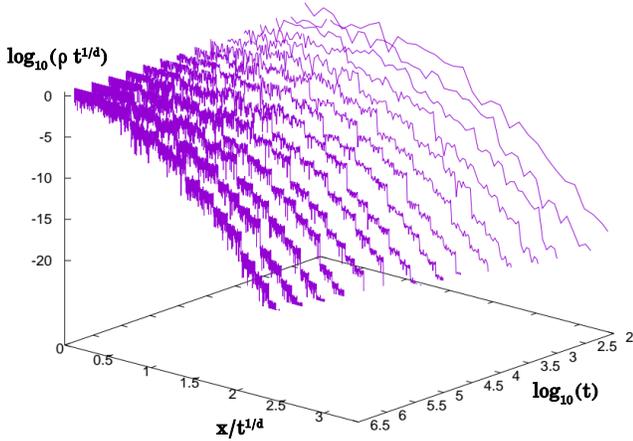}\hfill{}

\vspace{-3cm}

\caption{\label{fig:UQWPDF}Plot of the PDF $\rho(x,t)$ as a function of pseudo-velocity
$x/t^{1/d_{w}}$, see Eq.~(\ref{eq:collapse}), for a range of times
$t$ after the initiation of the walk at $x=0$. (Only $x\geq0$ is
shown.) In these simulations, we set $\epsilon=\frac{1}{2}$ so that
$d_{w}=1.661$, according to Eq.~(\ref{eq:dwQ}). }
\end{figure}

While it is easy to simulate QW for any $\epsilon$, it remains a
challenge to interpret the data, in particular, to test Eq.~(\ref{eq:dwQ}).
Unlike for RW, the deterministic, unitary evolution of QW does not
provide for much of a stochastic variation over which repeated walks
could be averaged. Additionally, sometimes dramatic changes in the
behavior of a walk occur especially near large barriers. Both of these
effects are on display in Fig.~\ref{fig:UQWPDF}, showing the PDF
of QW with $\epsilon=\frac{1}{2}$, re-scaled according to Eq.~(\ref{eq:collapse}).
Large spatial fluctuations, and ever more pronounced jumps near larger
barriers, make it difficult to collapse the data with any accuracy.
However, asymptotically such a collapse with $d_{w}=1.661$, as Eq.~(\ref{eq:dwQ})
provides, seems plausible.

\subsection{Quantum walk with absorbing walls\label{subsec:QW-Abso}}

Unlike for the classical case discussed in Sec.~\ref{subsec:RW-Abso},
where the total absorption amounts conveniently to taking a \emph{local}
limit for $z\to1$ on the Laplace transform of the site amplitude,
evaluating the adsorption in QW is surprisingly involved, in comparison.
Due to unitarity, the adsorption sums up the square modulus of site
amplitudes, which correspond to contour integrals over the entire
unit circle in the complex $z-$plane for the modulus of their Laplace
transforms, 
\begin{equation}
F_{0}=\sum_{t=0}^{\infty}\left|\psi_{0,t}\right|^{2}=\oint\frac{dz}{2\pi iz}\left|\overline{\psi}_{0}\left(z\right)\right|^{2},\label{eq:Pplus}
\end{equation}
Such an integral is readily, albeit strenuously, evaluated for the
simple case of homogeneous QW ($\epsilon=1$), using the non-linear
Riemann-Lebesgue lemma\ \cite{Ambainis01}. However, for $\epsilon<1$
we have only the local asymptotic evaluation of the RG recursions
in Eq.~(\ref{eq:1dQWrecursions}) available. While, for instance,
$\overline{\psi}_{0}\left(z\right)$ in Eq.~(\ref{eq:Pplus}) is
a functional of the hopping operators, inserting their asymptotic
form for a \emph{local} expansion of the integral does not appear
to be sufficient to obtain the absorption.

As mentioned above, resorting instead to a direct simulation of QW
shows that it is difficult to extract the scaling for moments of the
walk to, say, verify the walk dimension in Eq.~(\ref{eq:dwQ}) with
any reasonable accuracy. The irregular pattern of reflecting barriers
leads to very noisy probability densities. However, it is quite easy
to convince oneself, starting from very small systems and progressively
doubling their size, that the total absorption remains exactly unity
throughout for even the smallest values of $\epsilon$. Thus, there
appears to be no localization in the interior of the system; all of
the weight of QW eventually reaches a wall!

\section{Conclusions\label{sec:Discussion}}

The ultra-walk provides an exactly solvable model of a walk, both
classical or quantum, with tunable spatial heterogeneity. For the
classical case, we reproduce previous results by alternative means,
using a $2^{{\rm nd}}-$order Markov process that closely resembles
the coined QW in form. For the discrete-time QW, we obtain entirely
new results over the entire range of heterogeneity, with walk dimensions
ranging from $d_{w}^{QW}=1$ to infinity. Thus, while RW is quite
robust against the introduction of those barriers, even the smallest
amount of inhomogeneity changes the asymptotic behavior of QW. However,
numerical verification of these results for QW is difficult to obtain,
due to the strong hierarchical nature. Focusing on the absorption
problem at walls that recede from the starting site with increasing
system size, we can at least ascertain, as in the classical case,
that there is no localization, even for the most extreme heterogeneity.
Future work will focus on the asymptotic evaluation of observables,
like the absorption in Eq.~(\ref{eq:Pplus}), that are defined via
a complex integration for QW. Absorption, transit and first passage
problems of that sort are fundamental to any transport problem, classical
or quantum \cite{VanKampen92,Redner01}.

\paragraph*{Acknowledgments:}

SB acknowledges financial support during the preparation of this manuscript
from the U. S. National Science Foundation through grant DMR-1207431
and from CNPq through the ``Ciência sem Fronteiras'' program, and
he thanks Renato Portugal, Stefan Falkner, and Shanshan Li for many
helpful discussions and LNCC for its hospitality.

\bibliographystyle{apsrev4-1}
\bibliography{/home/stb/Boettcher}

\end{document}